\documentclass[amsmath,epsf,eclepsf,epsfig,subfigure,preprint,aps,graphicx,amssymb]{revtex4}
\usepackage[dvips]{graphicx}
\usepackage{color}
\usepackage{amssymb}
\preprint{HUPD0502}
\begin{document}
\title{\Large\bf Cosmological Family Asymmetry and
 CP violation}
\author{
T. Fujihara$^{(a)}$
\footnote{fujihara@theo.phys.sci.hiroshima-u.ac.jp} S.
Kaneko$^{(b)}$ \footnote{sotoru@phys.ocha.ac.jp} S. Kang $^{(c)}$
\footnote{skkang@phya.snu.ac.kr}
\\
D. Kimura $^{(a)}$
\footnote{kimura@theo.phys.sci.hiroshima-u.ac.jp}
T. Morozumi$^{(a)}$
\footnote{morozumi@hiroshima-u.ac.jp}
M. Tanimoto $^{(d)}$
\footnote{tanimoto@muse.sc.niigata-u.ac.jp}
}
\address{
$^{(a)}$ Graduate School of Science, Hiroshima University,
 Higashi-Hiroshima, Japan, 739-8536}
\address{
$^{(b)}$
Department of Physics,
 Ochanomizu University, Tokyo, Japan 112-8610}
\address{
$^{(c)}$ Seoul National University, Seoul, Korea 151-734}
\address{
$^{(d)}$
Department of Physics, Niigata University, Niigata, Japan 950-2181}
\begin{abstract}
We discuss how the cosmological baryon asymmetry
can be achieved by the lepton family asymmetries of heavy Majorana
neutrino decays and they are related to CP violation in neutrino
oscillation, in the minimal seesaw model with two heavy Majorana
neutrinos.  We derive the most general 
formula for CP violation in neutrino oscillation
in terms of the heavy Majorana masses and Yukawa mass term. 
It is shown that the formula is very useful to classify several
models in which $e-$, $\mu-$ and $\tau-$leptogenesis can 
be separately realized and to see how they are connected
with low energy CP violaton. To make the models predictive,
we take texture with two zeros in the Dirac
neutrino Yukawa matrix. In particular, we find some interesting
cases in which CP violation in neutrino oscillation can happen
while lepton family asymmetries do not exist at all.  On the contrary,
we can find $e-$,
$\mu-$ and
$\tau-$leptogenesis scenarios in which the
cosmological CP
violation and low energy CP violation measurable via
neutrino
oscillations are very closely related to each other.
By
determining the allowed ranges of the parameters in
the models,
we predict
the sizes of CP violation in neutrino oscillation and
$|V_{e3}^{MNS}|$. Finally, the leptonic unitarity
triangles are reconstructed.
\end{abstract}
\maketitle
\newpage
\newcommand{\dis}[1]{\begin{equation}\begin{split}#1\end{split}\end{equation}}
\newcommand{\beqa}[1]{\begin{eqnarray}#1\end{eqnarray}}
\def\Im{\rm{Im}}
\def\be{\begin{equation}}
\def\ee{\end{equation}}
\def\bea{\begin{eqnarray}}
\def\eea{\end{eqnarray}}
\def\ben{\begin{enumerate}}
\def\een{\end{enumerate}}
\def\nn{\nonumber}
\def\lsl{ l \hspace{-0.45 em}/}
\def\ksl{ k \hspace{-0.45 em}/}
\def\qsl{ q \hspace{-0.45 em}/}
\def\psl{ p \hspace{-0.45 em}/}
\def\ppsl{ p' \hspace{-0.70 em}/}
\def\dsl{ \partial \hspace{-0.45 em}/}
\def\Dsl{ D \hspace{-0.55 em}/}
\def\matrix{ \left(\begin{array} \end{array} \right) }
\def\ma{m_A}
\def\mf{m_f}
\def\mz{m_Z}
\def\mw{m_W}
\def\ml{m_l}
\def\ms{m_S}
\def\dag{\dagger}
\def\hf{\textstyle{\frac12~}}
\def\hff{\textstyle{\frac13~}}
\def\hfg{\textstyle{\frac23~}}
\def\Arg{{\rm Arg}}
\def\Im{{\rm Im}}
\def\Re{{\rm Re}}
\newcounter{ichi}
\setcounter{ichi}{1}
\newcounter{ni}
\setcounter{ni}{2}
\newcounter{san}
\setcounter{san}{3}
\newcounter{yon}
\setcounter{yon}{4}
\baselineskip 0.7cm
\bibliographystyle{plain}
\thispagestyle{empty}
\section{Introduction}
 CP violations in the neutrino seesaw models
have recently attracted much attention  because the measurements
of CP violation via neutrino oscillation are being planned in
future experiments and there may exist a connection between the
low energy neutrino CP violation and the matter and anti-matter
asymmetry of the universe through the leptogenesis scenario in the
seesaw models \cite{FY}. In contrast to the quark sector, since
the number of independent CP violating phases in the neutrino
seesaw models is more than one\cite{EMOP}, it is not
straightforward to discriminate the CP violating phases
contributing to the leptogenesis from the low energy
experiments \cite{BMNR}. One can show that the CP violation phases 
at high energy can contribute to the low energy
effective Majorana mass matrix 
 and thus they may be concerned with
a CP violating phase called $\delta$ in the standard
parametrization of MNS matrix, which is measurable from CP
violation in neutrino oscillation.  One might think
that non-zero $\delta$ may play a role in CP violation for
leptogenesis in the neutrino seesaw models. However, this is not
always the case, because several independent CP phases contribute
to both the leptogenesis CP violation at high energy and CP
violation of neutrino oscillation at low energy. There is the case
in which while at low energy the total effect of many CP phases
are cancelled but at high energy cosmological CP violation
remains. There is the opposite case in which the cosmological CP
violation vanishes while CP violation at low energy is non-zero.
Considering the situation, it is important to study CP violation
phenomena as much as possible both at high energy and low energy.
In the previous work \cite{EMX}, it was shown that the lepton
family asymmetries $Y_e$,$Y_{\mu}$ and $Y_{\tau}$ which are
generated by heavy Majorana neutrinos decays are sensitive to one
of the many CP violating phases. Though the total lepton asymmetry
$Y=Y_e+Y_{\mu}+Y_{\tau}$ remains as a constant, flavor composition
of the asymmetries $Y_e:Y_{\mu}:Y_{\tau}$ can vary by changing the
phase. As a particulary interesting case, the amount and the sign
of each lepton family asymmetry  $Y_i$ can be very different from
the total lepton asymmetry as $|Y| \ll |Y_{\mu}|,|Y_{\tau}|$. One
can also find the case \cite{EMX}, the lepton asymmetry $Y$ could
be dominated by a particular lepton family asymmetry as $Y \sim
Y_{\mu}$ or $Y \sim Y_{\tau}$.
 If this is the case, it indicates
the interesting scenario of baryogenesis that the matter in
the present universe
was originated by the second or the third family of leptons.
Interestingly, the models proposed in \cite{FGY}
correspond to the $\mu$ or $\tau$ family number dominant
leptogenesis scenarios.
In this work, we study how such scenario can be probed by
low energy flavor violating processes such as neutrino oscillations.
The paper is organized as follows.
In section \Roman{ni}, we
study how CP violating phases are related to
lepton family asymmetries.
 The reason why, in general, the family asymmetries can be
different from the total lepton number asymmetry
is shown  in  a comprehensive way.
Then  we show how they have some impact on
the CP violation in
the neutrino mixings by deriving the formula for
low energy CP violation neutrino mixings in terms of
the fundamental parameters for the minimal seesaw model.
The analytical formulae for MNS matrix are given both for
normal and inverted cases.
In section \Roman{san}, we focus on the textures with two zeros in Yukawa
mass terms. By using the mixing angles and mass squared diffrences
determined
by neutrino oscillation experiments,
we determine the parameters of the models and make prediction
on $|V_{e3}^{MNS}|$ and CP violation in neutrino oscillation.
Based on this numerical fit, we reconstruct the leptonic unitarity
triangles.  Section 
\Roman{yon}
is devoted to summary and discussion.
\section{CP violation related to the lepton family asymmetry }
We start with the lepton family asymmetries generated from heavy
Majorana neutrino decays, which are defined by \cite{EMX}, \bea
\epsilon_i^k=\frac{\Gamma[N^k \rightarrow l_i^- \phi^+]-\Gamma[N^k
\rightarrow l_i^+ \phi^-]} {\Gamma[N^k \rightarrow l_i^-
\phi^+]+\Gamma[N^k \rightarrow l_i^+ \phi^-]}, \label{familya}
\eea where  $i=(e, \mu ,\tau)$ and $N^k$ denotes $k-$ th heavy
Majorana neutrino. The total lepton number asymmetry from $N^k$ is
\cite{FY}, \bea \epsilon^k= \sum_{i=e,\mu,\tau} \epsilon_i^k {\rm
Br} (N^k \rightarrow {l_i}^{\pm} \phi^{\mp}), \label{totala} \eea
where ${\rm Br}$ denotes the tree level branching fraction. For
our purpose, let us focus on the seesaw model with two heavy
Majorana neutrinos
\cite{FGY}\cite{Endoh}\cite{GuoXing}\cite{MeiXing},
\newcommand{\ov}{\overline}
\newcommand{\f}{\frac}
\begin{eqnarray}
  {\cal L}_{m}=-y_{\nu}^{ik}\ov{L_i}N_{k}\tilde{\phi}
              -y_{l}^{i} \ov{L_i} l_{R_i} {\phi}
-\f{1}{2}\ov{N_{k}}^{c} M_{k} N_{k}+h.c., \label{lagrangian}
\end{eqnarray}
where $i=e,\mu,\tau$ and $k=1,2$. $L_i, l_R, \phi$ are $SU(2)$
lepton doublet fields, charged lepton singlet fields and Higgs
scalar, respectively. Here we take a basis in which both charged
lepton and singlet Majorana neutrino mass matrices are real
and diagonal. In this basis, the lepton family asymmetries given in
Eq.(\ref{familya}) can be written as \cite{EMX}, \bea
\epsilon_i^k&=&\frac{1}{8\pi} \sum\limits_{k'\neq
k}\left[I(x_{k'k}) \frac{\Im[(y_\nu^\dag
y_\nu)_{kk'}(y_\nu)^*_{ik}(y_\nu)_{ik'}]}
{|(y_\nu)_{ik}|^2}\right.\nn\\
&&\left.+\frac{1}{1-x_{k'k}}
\frac{\Im[(y_\nu^\dag y_\nu)_{k'k}(y_\nu)^*_{ik}(y_\nu)_
{ik'}]}{|(y_\nu)_{ik}|^2}\right],
\label{epsilon}
\eea
where $x_{k'k}=\frac{M_{k'}^2}{M^2_{k}}$ and $I(x)$
is given as \cite{FY}\cite{Covi},
\bea
I(x)&=&\sqrt{x}\left[1+\frac{1}{1-x}+(1+x)\ln\frac{x}{1+x}\right]\nn\\
     &=&\left\{\begin{array}{ll}
        -\frac{3}{2}x^{-1/2}& \ \ \ {\rm for} \ \ x\gg 1,\\
        -2x^{3/2}           & \ \ \ {\rm for} \ \ x\ll 1.
              \end{array}\right.
\label{Iformulae}
\eea
It is convenient to write $3 \times 2$ matrix Dirac mass
$m_D= y_{\nu} \frac{v}{\sqrt{2}}$ as,
\bea
m_D=
\left( \begin{array}{cc}
{\bf m_{D 1}}, & {\bf m_{D 2}} \end{array} \right)
=\left( \begin{array}{cc}
           m_{D e1} & m_{D e2} \\
           m_{D \mu 1} & m_{D \mu 2} \\
           m_{D \tau 1} & m_{D \tau 2} \end{array} \right)
= \left(\begin{array}{cc}
{\bf u_1} ,{\bf u_2} \end{array}\right)
\left( \begin{array}{cc} m_{D1}& 0 \\
                           0 & m_{D2} \end{array} \right),
\label{mD} \eea where two unit vectors are introduced, \bea {\bf
u_i}=\frac{\bf m_{Di}}{m_{Di}}, \label{unit} \eea with
$m_{Di}=|{\bf m_{Di}}|$. Without loss of generality, we can take
${\bf u_1}$ and ${\bf u_2}$ to be real and complex, respectively.
Then, three CP
violating phases correspond to arg($u_{2 i}$) ($i= e$, $ \mu $, and
$\tau$). With the definitions, one can write, \bea
&& {\rm Br}(N^k \rightarrow {l_i}^{\mp} \phi^{\pm})=|u_{ik}|^2, \nn \\
&& \epsilon_i^1 {\rm Br}(N^1 \rightarrow {l_i}^{\mp} \phi^{\pm})=\nn \\
&& \frac{(m_{D2})^2}{ 4 \pi v^2} \left( I\left(\frac{M_2^2}{M_1^2}\right)
\Im [ ({\bf u_1^{\dagger} \cdot u_2}) u_{i1}^{\ast} u_{i2} ]
+\frac{M_1^2}{M_1^2-M_2^2}
\Im[({\bf u_1^{\dagger} \cdot u_2})^{\ast} u_{i1}^{\ast} u_{i2} ]
\right), \nn \\
&& \epsilon_i^2 {\rm Br}(N^2 \rightarrow {l_i}^{\mp} \phi^{\pm})= \nn \\
&& -\frac{(m_{D1})^2}{ 4 \pi v^2} \left( I\left(\frac{M_1^2}{M_2^2}\right)
\Im [ ({\bf u_1^{\dagger} \cdot u_2}) u_{i1}^{\ast} u_{i2} ]
+\frac{M_2^2}{M_2^2-M_1^2}
\Im[({\bf u_1^{\dagger} \cdot u_2})^{\ast} u_{i1}^{\ast} u_{i2} ]
\right). \nn \\
\label{epsilon2} \eea
It is interesting to note that the lepton
family asymmetries are related to the following combinations of
Yukawa terms, \bea
&& A_{12}^e=({\bf u_1^{\dagger} \cdot u_2}) {u}_{e1}^{\ast} {u}_{e2}, \nn \\
&& A_{12}^{\mu}=({\bf u_1^{\dagger} \cdot
u_2}) {u}_{\mu 1}^{\ast} {u}_{\mu 2},\nn \\
&& A_{12}^{\tau}=({\bf u_1^{\dagger} \cdot
u_2}) {u}_{\tau 1}^{\ast} {u}_{\tau 2},
\label{A}
\eea
\bea
&& B_{12}^e= ({\bf u_1^{\dagger} \cdot u_2})^{\ast}
 {u}_{e1}^{\ast} {u}_{e2},\nn \\
&& B_{12}^\mu=({\bf u_1^{\dagger} \cdot
u_2})^{\ast} {u}_{\mu 1}^{\ast} {u}_{\mu 2},\nn \\
&& B_{12}^{\tau}=({\bf u_1^{\dagger} \cdot u_2})^{\ast} {u}_{\tau
1}^{\ast} {u}_{\tau 2}, \label{B} \eea 
where $ A_{12}^i=B_{12}^i \exp(2 i \gamma) $ 
with $\gamma ={\rm arg}({\bf u_1^\dagger \cdot u_2})$. 
In addition, $A_{12}$ and $B_{12}$
satisfy the following sum rules, \bea
A_{12}^e+ A_{12}^{\mu}+A_{12}^{\tau}=({\bf u_1^{\dagger} \cdot u_2})^2, \nn \\
B_{12}^e+B_{12}^{\mu}+B_{12}^{\tau}=|({\bf u_1^{\dagger} \cdot u_2})|^2.
\eea
The relations are shown in Fig.1, where $\gamma=\frac{\pi}{4}$
is taken.
They are  quadrangles in complex plane.
The imaginary part of $A$ is related to CP asymmetry of
leptogenesis.
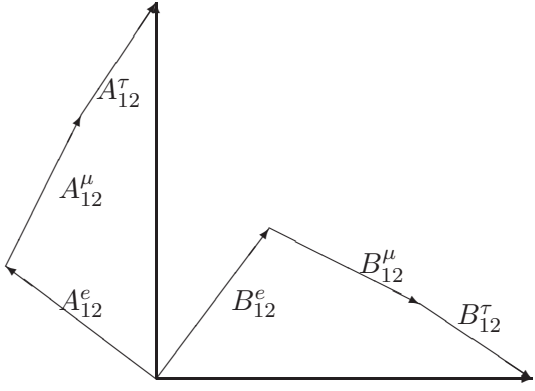
\begin{figure}
\setlength{\unitlength}{1mm}
\begin{picture}(100,40)(-30,0)
\put(3,5){\makebox(20,10){$B^e_{12}$}}
\put(20,10){\makebox(20,10){$B^\mu_{12}$}}
\put(33,3){\makebox(20,10){$B^\tau_{12}$}}
\put(-20,5){\makebox(20,10){$A^e_{12}$}}
\put(-20,20){\makebox(20,10){$A^\mu_{12}$}}
\put(-15,33){\makebox(20,10){$A^\tau_{12}$}}
\put(0,0){\vector(1,0){50}}
\put(0,0){\vector(3,4){15}}
\put(15,20){\vector(2,-1){20}}
\put(35,10){\vector(3,-2){15}}
\put(0,0){\vector(0,1){50}}
\put(0,0){\vector(-4,3){20}}
\put(-20,15){\vector(1,2){10}}
\put(-10,35){\vector(2,3){10}}
\end{picture}
\caption{Schematic view of quadrangles.}
\label{quad}
\end{figure}
The ratios of lepton family asymmetry to the total lepton
asymmetry are written as; \bea \frac{\epsilon^1_i}{\epsilon^1}&=&
\frac{I(\frac{M_2^2}{M_1^2}) \Im A_{12}^i +
\frac{M_1^2}{M_1^2-M_2^2} \Im B_{12}^i} {I(\frac{M_2^2}{M_1^2})
(\Im A_{12}^{e} + \Im A_{12}^{\mu} +
\Im A_{12}^{\tau})}, \nn \\
\frac{\epsilon^2_i}{\epsilon^2}&= &
\frac{I(\frac{M_1^2}{M_2^2}) \Im A_{21}^i + \frac{M_2^2}{M_2^2-M_1^2}
\Im B_{21}^i}
{I(\frac{M_1^2}{M_2^2}) (\Im A_{21}^{e} + \Im A_{21}^{\mu} +
\Im A_{21}^{\tau})}.
\eea
In the model with two heavy Majorana neutrinos $N_1$ and $N_2$
 with large hierarchical mass, e.g.,
$M_1 \ll M_2$, the family asymmetries from the lightest heavy
Majorana neutrinos decay are approximately given as, \bea
\frac{\epsilon^1_e}{\epsilon^1} \approx && \frac{\Im A_{12}^e}{\Im (A_{12}^{e}+A_{12}^{\mu}+A_{12}^{\tau})}, \nn \\
\frac{\epsilon^1_\mu}{\epsilon^1}\approx && \frac{\Im A_{12}^{\mu}}{\Im (A_{12}^{e}+A_{12}^{\mu}+A_{12}^{\tau})}, \nn \\
\frac{\epsilon^1_\tau}{\epsilon^1}\approx && \frac{\Im A_{12}^{\tau}}{\Im (A_{12}^{e}+A_{12}^{\mu}+A_{12}^{\tau})}.\nn \\
\label{approx}
\eea
Therefore one-family dominant leptogenesis can be realized
when the quadrangle is replaced by a line which
is determined by one of $A^{e}_{12},A^{\mu}_{12}$ and
$A^{\tau}_{12}$ with a non-trivial CP violating phase.
 If this is the case, the imaginary parts
of $A^{e}_{12}$, $A^{\mu}_{12}$ and $A^{\tau}_{12}$
are related to e-leptogenesis, $\mu$-leptogenesis and $\tau$-leptogenesis,
respectively. We also note that the imaginary part of $\sum_a A^{a}$
can be smaller
than the imaginary part of $A^{a}$. If this is the case, each family asymmetry is much larger
than the total lepton asymmetry.
Now let us discuss how the family asymmetry is related to the
CP violation in neutrino oscillations,
\bea
P(\nu_{\mu} \to \nu_{e})-P(\bar{\nu_{\mu}} \to \bar{\nu_{e}})
= 4 J \left( \sin \frac{\Delta m_{12}^2 L}{2 E}+
\sin \frac{\Delta m_{23}^2 L}{2 E}+
\sin \frac{\Delta m_{31}^2 L}{2 E} \right),
\eea
where $J$ is Jarlskog invariant \cite{jarls} defined as,
\bea
J={\rm Im}\left(V^{MNS}_{e1} V^{MNS \ast}_{\mu 1} V^{MNS \ast}_{e2}
 V^{MNS}_{\mu 2}\right).
\eea In the basis where the charged lepton mass matrix is
diagonal, $J$ is related to the following quantity \cite{BMNR},
\bea \Delta ={\rm Im} \left((m_{eff}m_{eff}^{\dagger})_{e \mu}
 (m_{eff} m_{eff}^{\dagger})_{\mu \tau}
(m_{eff} m_{eff}^{\dagger})_{\tau e} \right),
\eea
where $m_{eff}=-m_D \frac{1}{M} m_D^T$, and the relation between
$J$ and $\Delta$ is given as,
\bea
J=\frac{\Delta}{(n_1^2-n_2^2)(n_2^2-n_3^2)(n_3^2-n_1^2)},
\label{J}
\eea
where $n_i^2$ are three mass eigenvalues of $m_{eff} m_{eff}^{\dagger}$.
To facilitate the calculation of $\Delta$,
we introduce three $ 2 \times 2$
hermitian matrices $H_e$,$H_{\mu}$ and $H_{\tau}$,
\bea
H_i=\left( \begin{array}{cc}
\frac{|m_{D i1}|^2}{M_1} & \frac{m_{D i1} m_{D i2}^{\ast}}{\sqrt{M_1 M_2}} \\
\frac{m_{D i1}^{\ast} m_{D i2}}{\sqrt{M_1 M_2}}  & \frac{|m_{D i2}|^2}{M_2}
\end{array} \right), \quad (i=e, \mu, \tau),
\label{He} \eea and $\Delta$ is obtained by simply taking trace of
the product of $H$s, \bea \Delta={\rm Im Tr} \left({H^{\ast} H_e
H^{\ast} H_{\mu} H^{\ast} H_{\tau}} \right), \label{Deltatrace}
\eea with $H=H_e+H_{\mu}+H_{\tau}$. The formula given in terms of
$2\times 2$ matrices $H$ is useful and can be generalized to the
seesaw model including any number ($n_M$) of heavy Majorana
neutrinos by replacing $2 \times 2$ matrices $H$ in Eq.(\ref{He})
by $n_M \times n_M$ matrices.  Eq.(\ref{Deltatrace}) shows that CP
violations in neutrino oscillation is related to the imaginary
part of $H_{e 12}$, $H_{\mu 12}$ and $H_{\tau 12}$. We introduce
the following parameters with mass dimension, \bea
X_i=\frac{m_{Di}^2}{M_i}, \quad (i=1,2). \label{Xi} \eea By
substituting  Eq.(\ref{He}) into Eq.(\ref{Deltatrace}), we obtain,
\bea
\Delta=&& \left(1- {\bf |u_1^{\dagger} \cdot u_2|^2} \right) \times \nn \\
&& \Biggl(X_1^4 X_2^2
\left({\rm Im}
[\left(u_{e1}^{\ast} u_{e2} u_{\mu 1} u_{\mu 2}^{\ast} \right)
|u_{\tau 1}|^2
+\left( u_{\mu1}^{\ast} u_{\mu2} u_{\tau 1} u_{\tau 2}^{\ast} \right)
|u_{e 1}|^2
+\left (u_{\tau 1}^{\ast} u_{\tau 2} u_{e 1} u_{e 2}^{\ast} \right)
|u_{\mu 1}|^2]  \right)
\nn \\
&&+X_1^3 X_2^3
\left({\rm Im}
[(u_{e1}^{\ast} u_{e2}) ({\bf u_1^{\dagger} \cdot u_2})
     (|u_{\tau 1} u_{\mu 2}|^2-|u_{\mu 1} u_{\tau 2}|^2) \right.\nn \\
&& \left. \qquad \qquad  \qquad \qquad +
(u_{\mu 1}^{\ast} u_{\mu 2})({\bf u_1^{\dagger} \cdot u_2})
      (|u_{e 1} u_{\tau 2}|^2-|u_{\tau 1} u_{e 2}|^2)
      \right. \nn \\
&& \left. \qquad \qquad  \qquad \qquad +
(u_{\tau 1}^{\ast} u_{\tau 2})({\bf u_1^{\dagger} \cdot u_2})
      (|u_{\mu 1} u_{e 2}|^2-|u_{e 1} u_{\mu 2}|^2)] \right) \nn \\
&& -X_1^2 X_2^4
 \left( {\rm Im}
[\left(u_{e1}^{\ast} u_{e2} u_{\mu 1} u_{\mu 2}^{\ast} \right)
|u_{\tau 2}|^2
+\left(u_{\mu1}^{\ast} u_{\mu2} u_{\tau 1} u_{\tau 2}^{\ast} \right)
|u_{e 2}|^2
+\left(u_{\tau 1}^{\ast} u_{\tau 2} u_{e 1} u_{e 2}^{\ast} \right)
|u_{\mu 2}|^2] \right) \Biggr). \nn \\
\label{generaldelta} \eea This is the most general formula to
express the low energy CP violation measurable via neutrino
oscillation in terms of the Majorana masses and the Yukawa terms
in the seesaw model and a main result of the paper. In the model
with two heavy Majorana neutrinos,
the same quantity is computed by within two zeros texture
models in \cite{FGY}.  For the most general case, $J$ is obtained
by using bi-unitary parametrization of $m_D$ \cite{Endoh}. It is
worthwhile to note that the terms proportional to $ X_1^3 X_2^3$
are related to the family asymmetries because they are
proportional to $\Im A_{12}^e, \Im A_{12}^{\mu}$ and $\Im
A_{12}^{\tau}$. However, the terms proportional to $X_1^4 X_2^2 $
and $X_1^2 X_2^4$ are not directly related to $\Im A^a$.
Now, let us study the following two interesting cases.\\
(1)
$ {\bf u_1^{\dagger} \cdot u_2}=0$.\\
 This corresponds to the case that there is
no leptogenesis and any family asymmetries are vanishing.
However, CP violation in neutrino oscillation can occur in this
case because $\Delta$ is not vanishing, \bea \Delta=X_1^2 X_2^2
(X_1^2-X_2^2) \Im \left(u_{\tau 1}^{\ast} u_{\tau 2} u_{e1}
u_{e2}^{\ast} \right). \eea (2) ${\bf u_1^{\dagger} \cdot
u_2}=u_{1a}^{\ast} u_{2a}
\quad (a=e,\mu,\tau)$. \\
Each case for $a$ corresponds to one family dominant leptogenesis,
such as e-leptogenesis, $\mu$-letogenesis or $\tau$-leptogenesis.
This implies that the lepton asymmetry is dominated by one
particular family asymmetry. In order to see how the scenarios of
leptogenesis are connected with the low energy CP violation
parametrized by $\Delta$, we consider the Dirac neutrino Yukawa
matrix containing two zeros which makes the scenarios more
predictable. In this class of the models, the light neutrino mass
matrix given by $m_{eff}$ can be
parametrized by five independent parameters. From the experimental
results on three mixing angles and two mass squared differences,
the five parameters including a CP phase are strongly constrained.
In table \ref{table1} and table \ref{table2}, we classify the
models with two zeros texture into type I and II depending on the
positions of the two zeros in the neutrino Dirac Yukawa
matrix. As one can see from table \ref{table1},  for type I
models, $\Delta$ is generally non-vanishing and proportional
to ${\Im} (u_{a1}^{\ast} u_{a2})^2$, which implies that there
exists a strong correlation between low energy CP violation and
leptogenesis. In contrast to the type I models, for the type II
models, the low energy CP violating parameter $ \Delta $ is
vanishing  and thus it is difficult to trace the origin of
cosmological family asymmetries from the measurement of the CP
violation in neutrino oscillation.
\begin{table}
\caption{Type I texture models and low energy CP violation}
\label{table1}
\begin{center}
\begin{tabular}{|c|c|c|}
\hline
Type &   & $\Delta$ \\ \hline
{\rm Type I (a)}
e-leptogenesis  &
$ \left(
\begin{array}{cc}
u_{e1} & u_{e2} \\
u_{\mu 1}    & 0\\
0 & u_{\tau 2} \\
\end{array} \right)$ &
$(1-|u_{e1} u_{e2}|^2)
X_1^3 X_2^3 {\rm Im} (u_{e1}^{\ast} u_{e2})^2 (-|u_{\tau2}|^2 |u_{\mu 1}|^2)$
\\ \hline
{\rm Type I(b)}
e-leptogenesis &
$ \left(
\begin{array}{cc}
u_{e1} & u_{e2} \\
0 & u_{\mu 2} \\
u_{\tau 1} & 0
\end{array}
\right)$ &
$(1-|u_{e1} u_{e2}|^2)
X_1^3 X_2^3 {\rm Im} (u_{e1}^{\ast} u_{e2})^2 |u_{\tau 1}|^2 |u_{\mu 2}|^2.
$ \\ \hline
{\rm Type I (a)} $\mu$ leptogenesis&
$\left(
\begin{array}{cc}
u_{e1} & 0 \\
u_{\mu 1}  & u_{\mu 2} \\
0 & u_{\tau 2} \\
\end{array} \right)$  &
$ (1-|u_{\mu 1} u_{\mu 2}|^2)
X_1^3 X_2^3 {\rm Im} (u_{\mu1}^{\ast} u_{\mu2})^2 (|u_{\tau 2}|^2 |u_{e 1}|^2)
$ \\ \hline
{\rm Type I (b)}
$\mu$ leptogenesis &
$
\left(
\begin{array}{cc}
0 & u_{e 2} \\
u_{\mu 1} & u_{\mu 2} \\
u_{\tau 1} & 0 \\
\end{array} \right)
$ &
$ (1-|u_{\mu 1} u_{\mu 2}|^2) X_1^3 X_2^3
{\rm Im} (u_{\mu1}^{\ast} u_{\mu2})^2 (-|u_{e 2}|^2 |u_{\tau 1}|^2)
$ \\ \hline
{\rm Type I (a) $\tau$ leptogenesis} &
$ \left(
\begin{array}{cc}
u_{e 1} & 0 \\
0 & u_{\mu 2} \\
u_{\tau 1} & u_{\tau 2} \\
\end{array} \right) $  & $ (1-|u_{\tau 1} u_{\tau 2}|^2)
X_1^3 X_2^3 {\rm Im} (u_{\tau1}^{\ast} u_{\tau2})^2
(-|u_{e 1}|^2 |u_{\mu 2}|^2) $  \\
\hline
{\rm Type I (b)} $\tau$ leptogenesis & $
\left(
\begin{array}{cc}
0 & u_{e2} \\
u_{\mu 1} & 0 \\
u_{\tau 1} & u_{\tau 2}
\end{array} \right) $ & $
(1-|u_{\tau 1} u_{\tau 2}|^2) X_1^3 X_2^3
{\rm Im} (u_{\tau 1}^{\ast} u_{\tau 2})^2 (|u_{e 2}|^2 |u_{\mu 1}|^2)
$ \\  \hline
\end{tabular}
\end{center}
\end{table}
\section{Neutrino mass spectrum and their mixings}
First we examine the neutrino mass spectrum. The eigenvalue
equation for $ m_{eff}$ is given by $ \rm{det} (m_{eff}
m_{eff}^{\dagger}-\lambda)=0$, where $\lambda$
denotes the eigenvalues for mass squared matrix and can be
determined by the following equations,
 \bea \lambda^3-\lambda^2 {\rm Tr} \left({m_D
\frac{1}{M} m_D^T m_D^{\ast} \frac{1}{M} m_D^{\dagger}} \right) +
\lambda \left(\frac{{\rm det} (m_D^{\dagger} m_D)}{M_1 M_2}
\right)^2=0. \eea Three mass eiegenvalues of $m_{eff}$ are related
with the MNS matrix through the following equation,
\bea
V^{MNS \dagger} m_{eff} V^{MNS \ast} = \left(\begin{array}{ccc}
                                    n_1 & 0 & 0 \\
                                    0 & n_2 & 0 \\
                                    0 & 0 & n_3 \\
                                    \end{array} \right).
\eea
We note that, in the minimal seesaw model with two heavy
Majorana neutrinos, there are one massless neutrino and two
massive neutrino whose masses are given by, \bea
n_{\pm}^2&=&\frac{X_1^2+X_2^2+ 2 X_1 X_2 Re.({\bf u_1^\dagger
\cdot u_2})^2}{2}
\nn \\
  &&\pm\frac{\sqrt{\left(X_1^2+X_2^2+ 2 X_1 X_2
   {\rm Re}.({\bf u_1^\dagger \cdot u_2})^2\right)^2
     - 4 X_1^2 X_2^2\left(1-|{\bf u_1^{\dagger} \cdot u_2}|^2\right)^2}}{2}.
     \label{masses}
\eea For the normal hierarchical case, the mass spectrum is given by,
\bea
n_1^2&=&0, \nn \\
n_2^2&=&\Delta m_{sol}^2=n_{-}^2, \nn \\
n_3^2&=&\Delta m_{atm}^2+\Delta m_{sol}^2=n_{+}^2, \label{Normal}
\eea and for the inverted mass hierarchical case \cite{GuoXing},
it is
\bea
n_1^2&=&\Delta m_{atm}^2-\Delta m_{sol}^2=n_{-}^2, \nn \\
n_2^2&=&\Delta m_{atm}^2=n_{+}^2, \nn \\
n_3^2&=&0. \label{Invert} \eea Now, let us consider how to obtain
the MNS matrix $V^{MNS}$. The diagonalization of $m_{eff}$
can be implemented  by two steps. First, we decouple a massless
state by rotating $m_{eff}$ with a unitary transformation $V$.
Then, the rotated mass matrix contains nontrivial two by two part
which is diagonalized by another unitary matrix $K$. The MNS
matrix is then given by their product as follows, \bea V^{MNS}=VK.
\eea In fact, the unitary matrix $V$ can be found from the
following relations:\\
for the normal hierachical case, denoting it as $V_N$,
 \bea
{V_N}^{\dagger} m_D& =\left(\begin{array}{cc}
            0 & 0 \\
            0 & \ast \\
            \ast & \ast \\
            \end{array} \right),
\label{VNiso} \eea and for the inverted hierarcical case, denoting it
as $V_I$  \bea V_I^{\dagger} m_D& =\left(\begin{array}{cc}
            \ast & \ast \\
            0& \ast \\
            0 & 0 \\
           \end{array} \right).
\label{VIiso}
\eea
Using the two unit vectors defined in Eq.(\ref{unit}),
the matrix $V_N $ and $V_I$ can be written as,
\bea
V_N=\left( \begin{array}{ccc}
 \frac{{\bf u_2^{\ast} \times u_1^{\ast}}}
         {\sqrt{1-{\bf |u_1^{\dagger} \cdot  u_2|^2}}}, &
  \frac{\bf u_2 - (u_1^{\dagger}\cdot u_2) u_1
 }{\sqrt{1-{\bf |u_1^{\dagger} \cdot  u_2|^2}}}, & {\bf u_1}
 \end{array} \right),
\label{VN}
\eea
\bea
V_I=\left( \begin{array}{ccc}
{\bf u_1}, &
   \frac{\bf u_2 - (u_1^{\dagger}\cdot u_2) u_1
 }{\sqrt{1-{\bf |u_1^{\dagger} \cdot  u_2|^2}}},
 & \frac{{\bf u_2^{\ast} \times u_1^{\ast}}}
         {\sqrt{1-{\bf |u_1^{\dagger} \cdot  u_2|^2}}}
 \end{array} \right).
\label{VI} \eea From Eq.(\ref{VNiso}) and Eq.(\ref{VIiso}), we
indeed see that  a massless state  is decoupled as, \bea
Z_N=V_N^{\dagger} m_{eff} V_N^{\ast}=\left(\begin{array}{ccc}
                                         0 & 0& 0 \\
                                         0 &Z_{N22} & Z_{N23} \\
                                         0 & Z_{N23} &Z_{N33} \\
                                         \end{array} \right),
\eea
where,
\bea
Z_{N22}&=&-X_2 \left(1-|{\bf u_1^{\dagger} \cdot u_2}|^2 \right),\nn \\
Z_{N33}&=&-\left(X_1 + X_2 ({\bf u_1^{\dagger}  \cdot u_2})^2 \right),
\nn \\
Z_{N23}&=&-X_2 \sqrt{1-|{\bf u_1^{\dagger} \cdot u_2}|^2}
\left({\bf u_1^{\dagger} \cdot u_2}\right).
\eea
For the inverted hierachical case,
\bea
V_I^{\dagger} m_{eff} V_I^{\ast}=\left(\begin{array}{ccc}
                                         Z_{I11} & Z_{I12}& 0 \\
                                         Z_{I12} &Z_{I22} & 0\\
                                         0 & 0 & 0 \\
                                         \end{array} \right),
\eea where, \bea Z_{I11}=Z_{N33}, \quad Z_{I12}=Z_{N23}, \quad
Z_{I22}=Z_{N22}. \eea Finally, the unitary matrix $K$ is obtained
from diagonalizing the $ 2 \times 2$ submatrix of $Z$. It can be
parametrized by an angle $\theta$ and two phases $\phi$ and
$\alpha$. The final form for $V^{MNS}$ for the normal hierarchical
case is prsented as, \bea
&&V_N^{MNS}=\nn \\
&&\left(\begin{array}{ccc}
 \frac{u_{\mu 2}^{\ast} u_{\tau 1}^{\ast} -u_{\tau 2}^{\ast} u_{\mu 1}^{\ast}}
 {\sqrt{1-|\bf u_1^{\dagger} \cdot u_2|^2}}
  &\frac{u_{e 2}-u_{e 1} \bf u_1^{\dagger} \cdot u_2}
  {\sqrt{1-|\bf u_1^{\dagger} \cdot u_2|^2}}
   & u_{e1} \\
  \frac{
  u_{\tau 2}^{\ast} u_{e 1}^{\ast}-u_{e 2}^{\ast} u_{\tau 1}^{\ast}
  }{\sqrt{1-|\bf u_1^{\dagger} \cdot u_2|^2}} &
  \frac{u_{\mu 2}-u_{\mu 1} \bf u_1^{\dagger} u_2}
  {\sqrt{1-|\bf u_1^{\dagger} \cdot u_2|^2}}
  &u_{\mu 1} \\
  \frac{u_{e 2}^{\ast} u_{\mu 1}^{\ast}-u_{\mu 2}^{\ast} u_{e 1}^{\ast}
  }{\sqrt{1-|\bf u_1^{\dagger} \cdot u_2|^2}}
  & \frac{u_{\tau 2}-u_{\tau 1} \bf u_1^{\dagger} u_2}
  {\sqrt{1-|\bf u_1^{\dagger} \cdot u_2|^2}}
  &  u_{\tau 1} \\
     \end{array} \right)
 \left( \begin{array}{ccc}
 1 & 0 & 0  \\
 0 & \cos\theta_N & \sin\theta_N e^{(-i \phi_N)} \\
 0 & -\sin\theta_N e^{i \phi_N}  & \cos\theta_N\\
      \end{array} \right)
 \left( \begin{array}{ccc}
 1 & 0 & 0  \\
 0 & e^{i \alpha_N}& 0\\
 0 & 0  & e^{-i \alpha_N}\\
      \end{array} \right),\nn \\
\label{VN}
\eea
where $\theta_N$, $\phi_N$ and $\alpha_N$  are given as,
\bea
\tan 2 \theta_N &=& \left(\frac{2 |Z_{N22}^{\ast} Z_{N23} + Z_{N23}^{\ast} Z_{N33}|}{|Z_{N33}|^2-|Z_{N22}|^2} \right), \nn \\
&=&\frac{2 X_2 \sqrt{1-|{\bf u_1^{\dagger} \cdot u_2}|^2}
|X_1 {\bf (u_1^{\dagger} \cdot u_2)^{\ast}}+ X_2
{\bf u_1^{\dagger} \cdot u_2}|}
{X_1^2+X_2^2(2 |{\bf u_1^{\dagger} \cdot u_2}|^2-1) +
2 X_1 X_2 {\rm Re.}({\bf u_1^{\dagger} \cdot u_2})^2}.
\nn \\
\phi_N&=&{\it arg}.(Z_{N22}^{\ast} Z_{N23} + Z_{N23}^{\ast} Z_{N33}),
\nn \\
&=&{\it arg}.(X_1 {\bf (u_1^{\dagger} \cdot u_2)^{\ast}}
 + X_2 {\bf (u_1^{\dagger} \cdot u_2})),
 \nn \\
2 \alpha_N&=
&{\it arg}.[ \cos^2\theta Z_{N22} + \sin^2\theta Z_{N33}
\exp(- 2 i \phi)- \nn \\
&& \sin 2 \theta Z_{N23} \exp(-i \phi)]. \label{theta} \eea The
mixing angle $ \theta_N$ can be unambiguously determined by
requiring the condition, $ \sin \theta_N \cos\theta_N \ge 0$, so
that the normal mass hierarchy ($n_2^2 \le n_3^2$) is maintained.
For the inverted hierarchical case, MNS matrix becomes, \bea
&& V_I^{MNS} \nn \\
&& =\left(\begin{array}{ccc}
 \frac{u_{e 2}-u_{e 1} \bf u_1^{\dagger} \cdot u_2}
  {\sqrt{1-|\bf u_1^{\dagger} \cdot u_2|^2}} & u_{e1}
   &
 \frac{u_{\mu 2}^{\ast} u_{\tau 1}^{\ast} -u_{\tau 2}^{\ast} u_{\mu 1}^{\ast}}
 {\sqrt{1-|\bf u_1^{\dagger} \cdot u_2|^2}}
    \\
  \frac{u_{\mu 2}-u_{\mu 1} \bf u_1^{\dagger} \cdot u_2}
  {\sqrt{1-|\bf u_1^{\dagger} \cdot u_2|^2}}
  & u_{\mu1} &
   \frac{
  u_{\tau 2}^{\ast} u_{e 1}^{\ast}-u_{e 2}^{\ast} u_{\tau 1}^{\ast}
  }{\sqrt{1-|\bf u_1^{\dagger} \cdot u_2|^2}}
  \\
  \frac{u_{\tau 2}-u_{\tau 1} \bf u_1^{\dagger} u_2}
  {\sqrt{1-|\bf u_1^{\dagger} \cdot u_2|^2}} & u_{\tau1} &
  \frac{u_{e 2}^{\ast} u_{\mu 1}^{\ast}-u_{\mu 2}^{\ast} u_{e 1}^{\ast}
  }{\sqrt{1-|\bf u_1^{\dagger} \cdot u_2|^2}}
     \end{array} \right)
 \left( \begin{array}{ccc}
 \cos\theta_I & \sin\theta_I e^{(-i \phi_I)} & 0\\
 -\sin\theta_I e^{i \phi_I}  & \cos\theta_I  & 0\\
 0 & 0 & 1
      \end{array} \right)
 \left( \begin{array}{ccc}
 e^{i \alpha_I} & 0 & 0  \\
 0 & e^{-i \alpha_I}& 0\\
 0 & 0  & 1 \\
      \end{array} \right),\nn \\
\label{VI} \eea where $\theta_I$ $\phi_I$ and $\alpha_I$ have the
same expressions as the normal hierarchical case given in terms of
$X_1, X_2,{\bf u_1}$ and ${\bf u_2} $. The condition, $\sin
\theta_I \cos \theta_I \ge 0$ ( $n_1^2 \le n_2^2$), is required
for inverted hierachical case. Having established how to construct
MNS matrix, we study the flavor mixings of two zeros texture models 
which are discussed in the previous section. We first study zero
of MNS matrix elements of type II models. The type II models
predict that one of the MNS matrix elements is zero. Because
experimental constraints allow $|V_{e3}^{MNS}|$ can be vanishing,
among type II models, only e-leptogenesis and inverted
hierarchycal case is allowed.
\begin{table}
\caption{Type II texture models and MNS matrix}
\label{table2}
\begin{center}
\begin{tabular}{|c|c|c|c|}
\hline
 type  &(a)$ \quad \quad \quad $ (b) &  $V^{MNS N}$  &  $V^{MNS I}$ \\ \hline
type II (e-leptogenesis) &
$ \left(\begin{array}{cc} u_{e1} & u_{e2} \\
                                         0 & u_{\mu 2} \\
                                         0 & u_{\tau 2}
                                         \end{array}\right)                     \quad \left(\begin{array}{cc} u_{e1} & u_{e2} \\
                                         u_{\mu1} & 0 \\
                                         u_{\tau1} & 0                                        \end{array}\right) $
 & $\left(\begin{array}{ccc} 0 & \ast & \ast \\
                                         \ast & \ast & \ast \\
                                         \ast & \ast & \ast
                                         \end{array}\right) $ &
$\left(\begin{array}{ccc} \ast & \ast & 0 \\
                           \ast & \ast & \ast \\
                           \ast & \ast & \ast \end{array} \right)$  \\
                           \hline
type II ($\mu$ -leptogenesis) &
$ \left(\begin{array}{cc} 0 & u_{e2} \\
                                         u_{\mu1} & u_{\mu2}\\
                                         0 & u_{\tau2}
                                         \end{array}\right)                     \quad \left(\begin{array}{cc} u_{e1} & 0 \\
                                         u_{\mu1} & u_{\mu2} \\
                                         u_{\tau1} & 0
                                         \end{array}\right) $   &
$\left(\begin{array}{ccc} \ast & \ast & \ast \\
                           0 & \ast & \ast \\
                           \ast & \ast & \ast \end{array} \right)$ &
 $\left(\begin{array}{ccc} \ast & \ast & \ast \\
                           \ast & \ast & 0 \\
                           \ast & \ast & \ast \end{array} \right)$ \\
                           \hline
type II ($\tau$ -leptogenesis) &
$ \left(\begin{array}{cc} 0 & u_{e2} \\
                          0 & u_{\mu2} \\
                          u_{\tau1} & u_{\tau2}
                                         \end{array} \right)
 \quad \left(\begin{array}{cc}
                                         u_{e1} & 0 \\
                                         u_{\mu1} & 0 \\
                                         u_{\tau1} & u_{\tau2}
                                         \end{array} \right) $   &
   $\left(\begin{array}{ccc} \ast & \ast & \ast \\
                                         \ast & \ast & \ast \\
                                         0 & \ast & \ast
                                         \end{array}\right) $  &
                     $\left(\begin{array}{ccc} \ast & \ast & \ast \\
                                               \ast & \ast & \ast \\
                                               \ast & \ast & 0
                                               \end{array} \right)$  \\
                                               \hline
                            \end{tabular}
                            \end{center}
\end{table}
About the type I models, in general, we do not have zero of the MNS
matrix elements. Therefore, we need to carry out the detailed numerical
study on the mixing angles, which will be presented in the next section.
\section{Numerical Analysis}
\subsection{Determination of parameters}
\begin{figure}\resizebox{10cm}{10cm}{
  \includegraphics{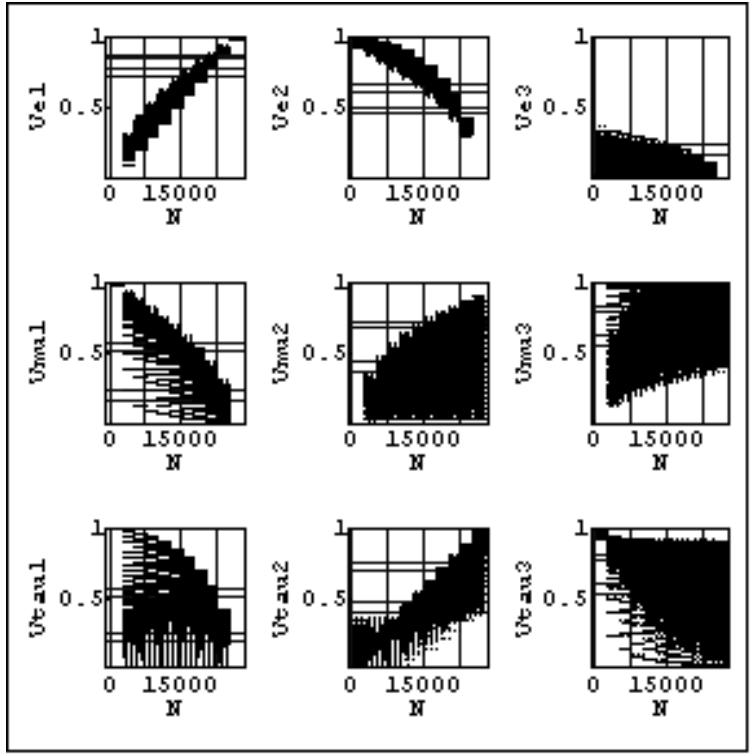}}
    \caption{
      $|V^{MNS}_{ij}|$
      for $\tau$ leptogenesis model type I(a) with normal hierarchy
      and $X_1 \le X_2$.
    }
  \label{MNS_tau1aN}
\end{figure}

From neutrino oscillation experiments, two mixing angles, the
upper bound on $|V_{e3}^{MNS}|$ and two neutrino mass squared
differences have been determined \cite{FT}\cite{GG}, which are
taken as inputs. Because in models with two zeros for
$m_D$, the effective low energy mass matrix $m_{eff}$ can be
presented in terms of five independent parameters including a CP
phase, all these parameters can be severely constrained from the
experimental results mentioned above. In this class of models, the
allowed ranges for $V_{e3}^{MNS}$ and Jarlskog invariant $J$
\cite{jarls} may be predicted. In this section, we determine the
allowed ranges for the parameters and predict $|V_{e3}^{MNS}|$ and
CP violation in neutrino oscillations $|J|$. Based on this
analysis, we can construct the possible forms of the unitarity
triangle
of leptonic sector.\\
We first show how two parameters $X_1$ and $X_2$ with mass
dimensions can be fixed by using $\Delta m^2$ and ${\bf
u_1^{\dagger} \cdot u_2}$ as inputs. Writing ${\bf u_1^{\dagger}
\cdot u_2}$ as, \bea {\bf u_1^{\dagger}\cdot u_2}=\cos \beta e^{i
\gamma}, \label{u1u2} \eea where $0 \le \cos \beta $ and $- \pi
\le \gamma \le \pi$, and using Eq.(\ref{masses}), we can write
$X_1+X_2$ and $|X_1-X_2|$ as, \bea X_1+X_2&=&\sqrt{n_+^2 + n_-^2+
2 n_+ n_- \cos 2 \gamma
+\frac{4 n_+ n_-}{\sin^2\beta} \sin^2 \gamma}, \nn \\
|X_1-X_2|&=& \sqrt{n_+^2 + n_-^2+ 2 n_+ n_- \cos 2 \gamma -\frac{4
n_+ n_-}{\sin^2 \beta} \cos^2 \gamma}. \label{X1X2} \eea Choosing
either $X_1 \le X_2$ or $X_1 \ge X_2$, we may write  $X_1$ and
$X_2$ in terms of $\beta$, $\gamma$ and neutrino masses. (See
Eq.(\ref{Normal}) and Eq.(\ref{Invert})). For numerical analysis,
we use $\Delta m^2_{sol.}=7.1 \times 10^{-5}$ (eV$^2$) and $\Delta
m^2_{atm.}=2.6 \times 10^{-3}$(eV$^2$). Here, we note that the
inputs $(\beta, \gamma, \Delta m^2_{sol.}, \Delta m^2_{atm.})$ are
sufficient for determining $ \sin \theta_{N,I} $ and $\phi_{N,I}$
in $K$ with the help of Eq.(\ref{theta}). We also note that $ \cos
\beta$ is bounded as, \bea \cos \beta \le
\frac{(n_+-n_-)}{\sqrt{n_+^2+n_-^2+2 n_+ n_- \cos 2 \gamma}}.
\label{cosbeta} \eea Next we illustrate how one can fit the models
with two zeros in $m_D$ by using the experimental results. As an
example, we take type I(a) $\tau$-leptogenesis model which is listed in
table \ref{table1}. In the model, $u_{\tau 1}$, $u_{e1}$ and
$u_{\mu2}$ can be taken to be real and positive and $u_{\tau 2}$
is a complex variable. From the $\tau$-leptogenesis assumption,
\bea u_{\tau 1} u_{\tau 2} = \cos \beta \exp(i \gamma). \eea 
By considering the range of the parameters;
$\cos \beta \le u_{\tau 1} \le 1, |\gamma| \le \pi, 
0 \le \beta \le \frac{\pi}{2}$,
one
can numerically generate $u_{\tau 1}$, $ \gamma $ and $\beta$ as, \bea u_{\tau
1} &=&\cos \beta + \frac{k}{N_k} (1-\cos \beta) \quad (k=0 \sim
N_k),\nn
\\
\gamma&=&- \pi + 2 \frac{(n_g-1) \pi }{N_g}, (n_g= 1 \sim N_{g}),\nn \\
\beta&=&\frac{(n_\beta-1) \pi}{2 N_\beta} \quad (n_\beta=1 \sim N_\beta),
\eea
where the number of divisions for each variable are taken
to be $N_\beta=N_g=50$ and $N_k=10$.
Then, we generate $(N_k+1) N_g N_\beta$ sets of $(\beta, \gamma, u_{\tau 1})$.
The other parameters in $ u_{ai} $ can be determined as,
\bea
u_{\tau 2}&=&\frac{\cos \beta \exp(i \gamma)}{u_{\tau 1}},\nn \\
u_{e 1}&=&\sqrt{1-|u_{\tau 1}|^2}, \nn \\
u_{\mu 2}&=&\sqrt{1-|u_{\tau 2}|^2}.
\label{us}
\eea
By fixing the parameters $(\beta, \gamma, u_{\tau 1})$ which is
equivalent to giving a set of three intergers $(n_{\beta}, n_{g}, k)$
, we can
generate all the elements of MNS matrix
through Eq.(\ref{us}), Eq.(\ref{X1X2}),
Eq.(\ref{u1u2}) and Eq.(\ref{VN}) $\sim$
Eq.(\ref{VI}).   To show how we determine the parameters
by taking into account of the experimental constraints,
it is convenient to
represent a set of the integers $(n_{\beta}, n_{g}, k)$
with an interger $N$ defined as,
\bea
N=k N_\beta N_g +(n_g-1) N_\beta + n_{\beta}.
\label{N}
\eea
For a given $N$, one can extract
a set of three integer
numbers $(n_\beta, n_g, k)$ as follows,
\bea
k&=&[\frac{N}{N_g N_\beta}], \nn \\
N^{\prime}&=&{\rm Mod}.[N, N_g N_\beta] ,\nn \\
n_g&=&[\frac{N^{\prime}}{N_\beta}]+1, \nn \\
n_{\beta}&=& {\rm Mod}.[N^{\prime}, N_\beta],
\label{integers}
\eea
where $[x]$ denotes the maximum integer which is not larger than $x$.
By taking $N$ in horizontal axis, we show the prediction
for the absolute values
of MNS matrix elements in vertical axis as shown in Fig.\ref{MNS_tau1aN}.
 A point of horizontal
axis corresponds to a set of parameters for ($ \beta, \gamma, u_{\tau 1}$).
We also show the experimentally allowed range for MNS matrix elements
both at
$90 \%$ confidence level and at $3 \sigma$ level
taken from \cite{GG}. One can find $N$ which leads to 
the MNS matrix elements consistent with
experiments. Then, we can determine $(n_\beta, n_g, k)$
by Eq.(\ref{integers}) and $(\beta, \gamma, u_{\tau 1})$
by Eq.(\ref{us}), respectively.
\begin{figure}\resizebox{10cm}{10cm}{
  \includegraphics{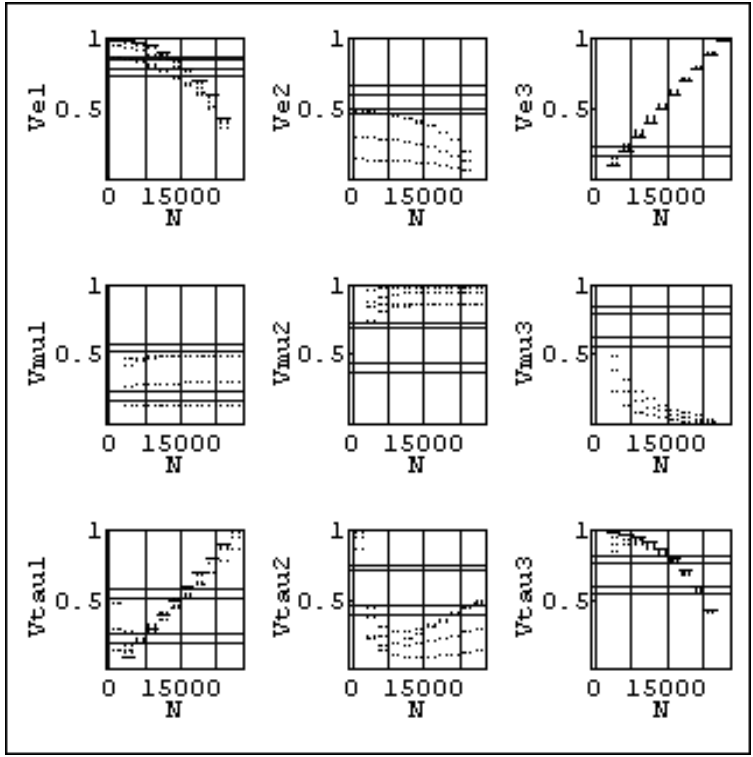}}
    \caption{
      $|V^{MNS}_{ij}|$
      for $\tau$ leptogenesis model type I(a) with inverted hierarchy
      and $X_1 \le X_2$.
    }
  \label{MNS_tau1aI}
\end{figure}
In Fig.\ref{MNS_tau1aI}, we show the fit for the inverted heirachical case.
By finding $N$ which reproduces the manignitude of five MNS matrix elements
simultaneously, we can determine the parameters of the model.
In this way, one can find $N$ which puts MNS matrix elements within
experimentally allowed range. In table \Roman{san},
we show our fit based on the experimental determination of the mixing angles
at $90 \%$ CL.
Only four types of textures are allowed and all the types correspond to normal
hierarchical case and either $\mu$ or $\tau$ leptogenesis case.
$|V^{MNS}_{e3}|$ is determined to be non-zero 
and the upper bound for CP violation $|J|$
is obtained. In table \Roman{yon}, we relax experimental constraints by using
$3 \sigma$ allowed range. In this case, more textures are allowed and
the allowed ranges are larger than previous case.
In addition to the
previous allowed textures,
the type II e-leptogenesis
(inverted hierarchical) case are allowed.
As for the type I $\mu$ and $\tau$ leptogenesis, the inverted hierachical
cases can be also fitted. Let us
summarize the fitted results for each texture
as follows.
\begin{table}
\begin{tabular}{|c|c|c|c|c|c|c|} \hline
type & $|V^{MNS}_{e1}|$   & $|V^{MNS}_{e2}|$    & $|V^{MNS}_{e3}|$
     & $|V^{MNS}_{\mu3}|$ & $|V^{MNS}_{\tau3}|$ & $|J|$ \\ \hline
     exp. ($90 \%$)
     &$ 0.79 \sim 0.86$   & $0.50 \sim 0.61$    & $0 \sim 0.16$
     & $ 0.63 \sim 0.79$  & $ 0.60 \sim 0.77$   &
\\  \hline
I(a) $\mu$ normal $X_1 \le X_2$
     & $0.79 \sim 0.86$   & $0.50 \sim 0.61$    & $0.058 \sim 0.11$
     & $ 0.63 \sim 0.79$  & $ 0.60 \sim 0.77$   & $0 \sim 0.023$
\\ \hline
I(b) $\mu$ normal $X_2 \le X_1$
     & $0.79 \sim 0.86$   & $ 0.50 \sim 0.61$   & $0.058 \sim 0.11$
     & $0.64 \sim 0.79$   & $0.61 \sim 0.77 $   & $0 \sim 0.024$
\\ \hline
 I(a) $\tau$ normal $X_1 \le X_2$
     & $0.79 \sim 0.86$   & $0.50 \sim 0.61$    & $0.054 \sim 0.10$
     & $0.63 \sim 0.79$   & $0.61 \sim 0.77$    & $0 \sim 0.022$
\\ \hline
I(b) $\tau$ normal $X_2 \le X_1$
     & $0.79 \sim 0.86$   & $0.50 \sim 0.61$    & $0.054\sim 0.10$
     & $0.63 \sim 0.79$   & $0.61 \sim 0.77$    & $0 \sim 0.022$
\\ \hline
\end{tabular}
\label{V90}
\caption{The predictions for $|V^{MNS}_{ij}|$ and $|J|$.  The magnitutes of the
$V^{MNS}_{ij}$ given in the second row correspond to experimental constraints
at $90 \% $ CL taken from \cite{GG}.}
\end{table}
\begin{table}
\begin{tabular}{|c|c|c|c|c|c|c|} \hline
type & $|V^{MNS}_{e1}|$   & $|V^{MNS}_{e2}|$    & $|V^{MNS}_{e3}|$
     &$|V^{MNS}_{\mu3}|$  & $|V^{MNS}_{\tau3}|$ & $|J|$
\\ \hline
exp.($3 \sigma $)
     & $ 0.73 \sim 0.88$  & $0.47 \sim 0.67$    & $0 \sim 0.23$
     & $ 0.56 \sim 0.84 $ & $ 0.54 \sim 0.82$   &
\\  \hline
I(a) $\mu$ normal $X_1 \le X_2$
     & $0.73 \sim 0.88$   & $0.47 \sim 0.67$    & $0.046 \sim 0.13 $
     & $ 0.57 \sim 0.83 $ & $ 0.54 \sim 0.82$   & $0 \sim 0.028$
\\ \hline
I(b) $\mu$ normal $X_2 \le X_1$
     & $0.73 \sim 0.88$   & $ 0.47 \sim 0.67$   & $ 0.047 \sim 0.13$
     & $0.57 \sim 0.83$   & $0.54 \sim 0.82 $   & $0 \sim 0.028$
\\ \hline
I(a) $\tau$ normal $X_1 \le X_2$
     & $0.73 \sim 0.88$   & $0.47 \sim 0.67$    & $0.044 \sim 0.13$
     & $0.56 \sim 0.84$   & $0.54 \sim 0.82$    & $0 \sim 0.027$
\\ \hline
I(b) $\tau$ normal $X_2 \le X_1$
     & $0.73 \sim 0.88$   & $0.47 \sim 0.67$    & $0.043\sim 0.12$
     & $0.56 \sim 0.84$   & $0.54 \sim 0.82 $   & $0 \sim 0.027$
\\ \hline
I(a) $\mu$ inverted $X_1 \le X_2$
     & $ 0.86 \sim 0.87$  & $0.48 \sim 0.49$    & $0.027 \sim 0.14$
     & $0.63 \sim 0.82$   & $0.56 \sim 0.77 $   & $0.0055 \sim 0.027$
\\ \hline
I(b) $\mu$ inverted $X_2 \le X_1$
    & $ 0.86 \sim 0.87$  & $0.48 \sim 0.49$    & $0.022 \sim 0.14$
    & $0.57 \sim 0.84$   & $0.54 \sim 0.82$    & $0.0044 \sim 0.028$
\\ \hline
I(a) $\tau$ inverted $X_1 \le X_2$
    & $ 0.86 \sim 0.87$  & $0.48 \sim 0.49$    & $0.027 \sim 0.13$
    & $0.59 \sim 0.84$   & $0.54 \sim 0.80 $   & $0.0055 \sim 0.026$
\\ \hline
I(b) $\tau$ inverted $X_2 \le X_1$
    & $ 0.86 \sim 0.87$ & $0.48 \sim 0.49$    & $0.021 \sim 0.13$
    & $0.57 \sim 0.84$  & $0.55 \sim 0.82 $   & $0.0039 \sim 0.027$
\\ \hline
II(a) e inverted $X_1 \le X_2$
     & $0.87$             & $0.49 \sim 0.50 $   & $0$
     & $0.57 \sim 0.84$   & $0.55 \sim 0.82$    & $0$
\\ \hline
II(b) e inverted  $X_2 \le X_1$
     & $0.87$             & $0.49 \sim 0.50 $   & $0$
     & $0.57 \sim 0.84$   & $0.55 \sim 0.82$    & $0$
\\ \hline
\end{tabular}
\label{Vsigma}
\caption{The predictions for $|V^{MNS}_{ij}|$ and $|J|$. The magnitutes of the
$V^{MNS}_{ij}$ given in the second row correspond to
experimental constraints
at $3 \sigma $ taken from \cite{GG}.}
\end{table}
\begin{itemize}
\item {Type II e-leptogenesis scenarios.
In this class of models,
because $|V^{MNS}_{e3}|=0$, CP violation in neutrino oscillation $J$
is vanishing in spite of non-zero $\gamma$.}
\item{Type I $\mu$ and $\tau$ leptogenesis for normal hierarchical case.
In this class of models, $V^{MNS}_{e3}$ is non-vanishing. About CP violation
phase, the allowed range of $|J|$ is from zero to some non-vanishing value.}
\item{Type I $\mu$ and $\tau$ leptogenesis for inverted hierarchical case.
In this class of models, both $V^{MNS}_{e3}$ and $|J|$
are non-vanishing.}
\end{itemize}
\subsection{$|V^{MNS}_{e3}|$ versus $|J|$}
\begin{figure}
\resizebox{10cm}{6cm}{
  \includegraphics{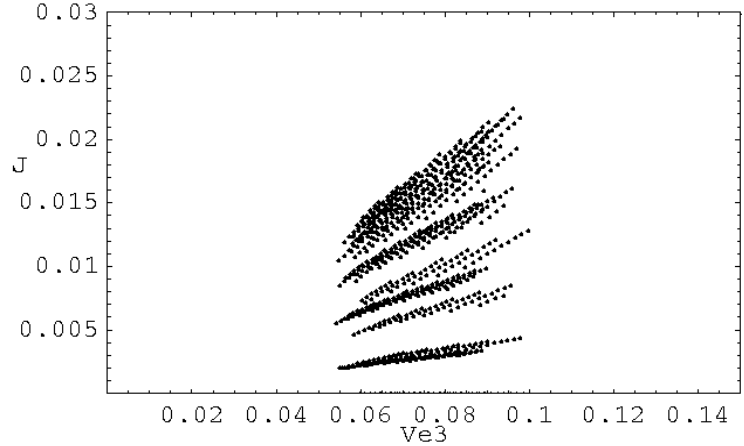}}
    \caption{
      $|J|$ and $|V^{MNS}_{e3}|$
      for $\tau$ leptogenesis model type I(a) with normal hierarchy
      and $X_1 \le X_2$. ($90 \%$)
    }
  \label{Ve3J1}
\end{figure}

\begin{figure}
\resizebox{10cm}{6cm}{
  \includegraphics{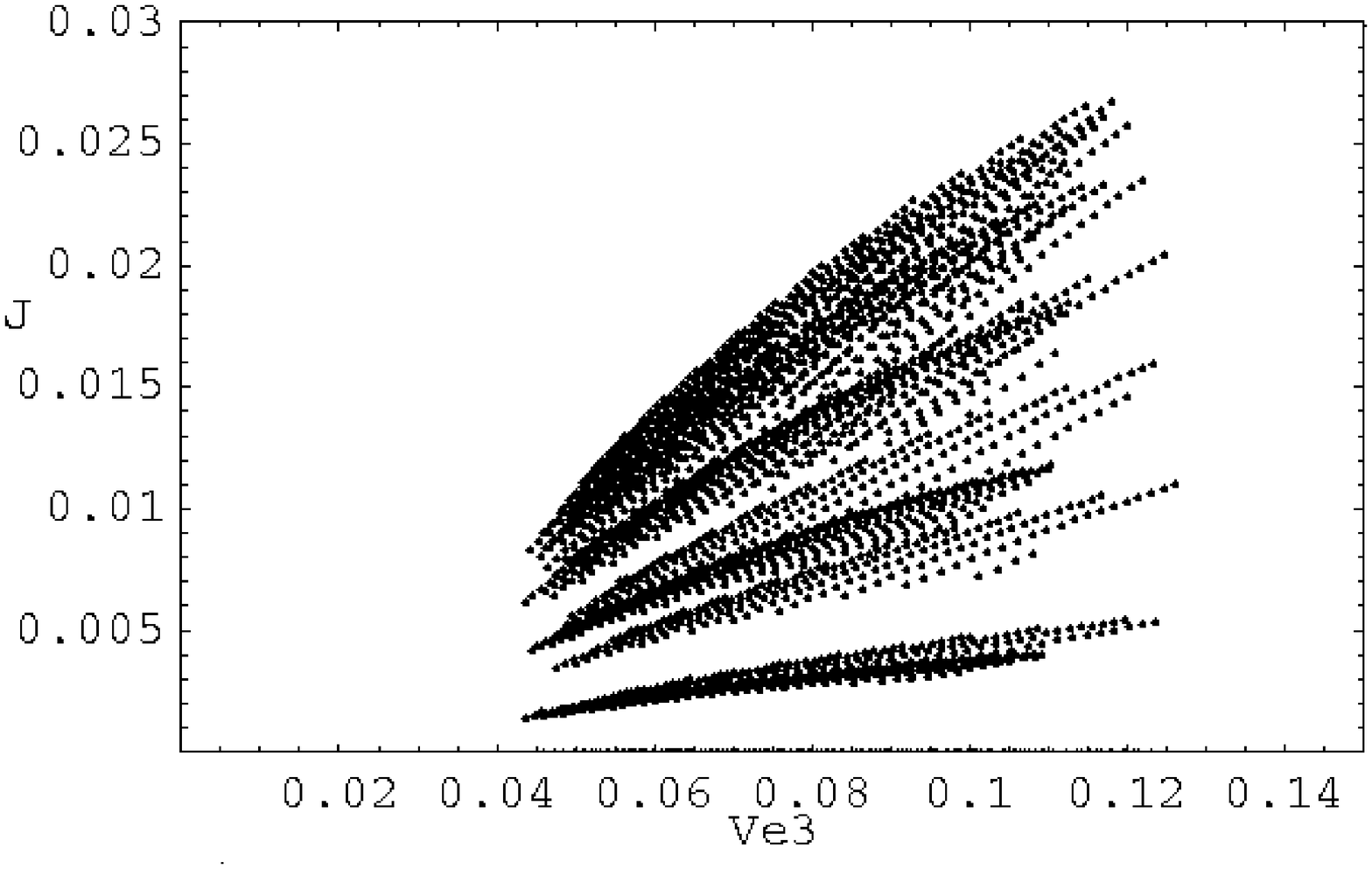}}
    \caption{
      $|J|$ and $|V^{MNS}_{e3}|$
      for $\tau$ leptogenesis model type I(a) with normal hierarchy
      and $X_1 \le X_2$. ($3 \sigma$)
    }
  \label{Ve3J2}
\end{figure}

\begin{figure}
\resizebox{10cm}{6cm}{
  \includegraphics{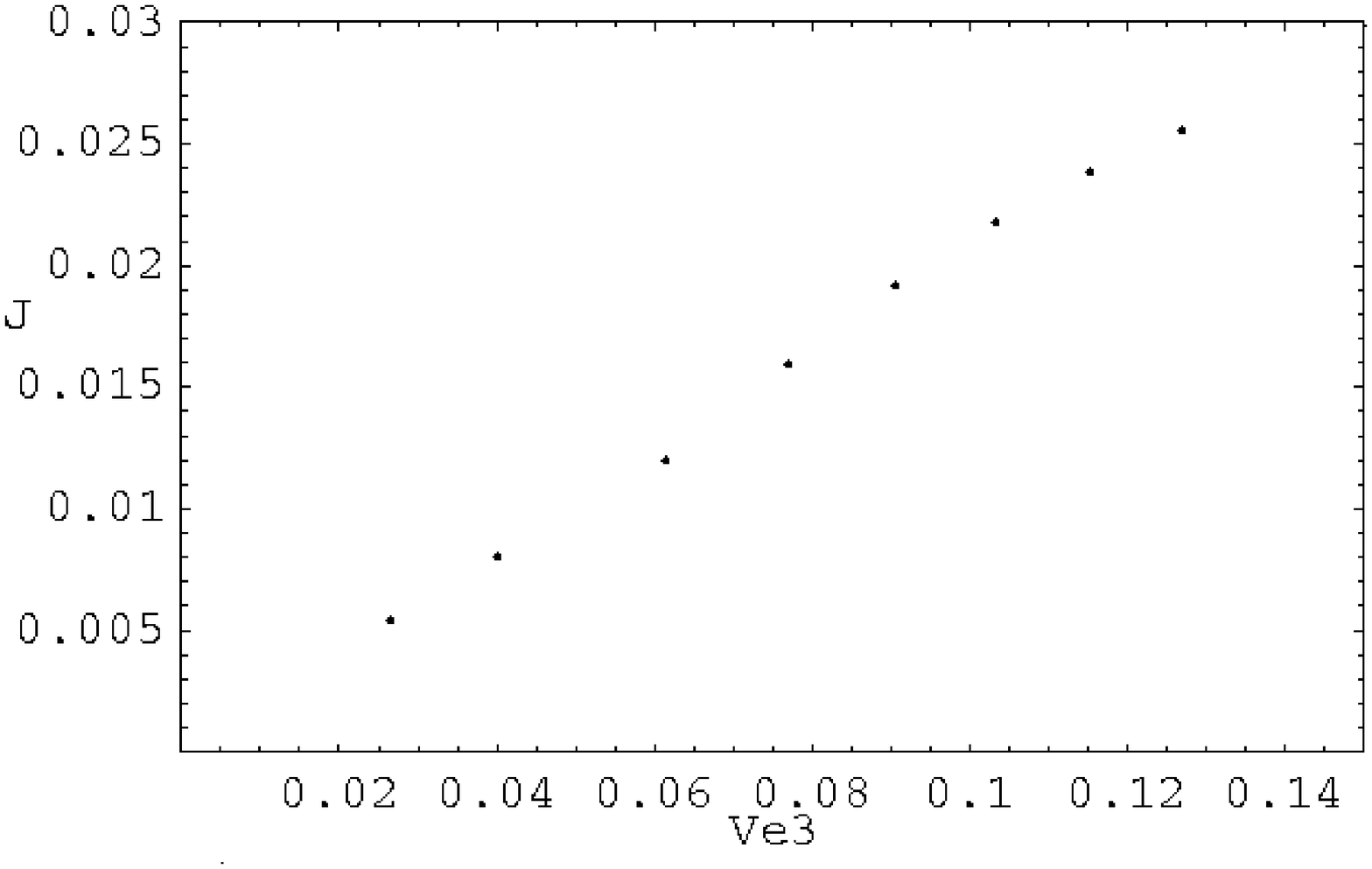}}
    \caption{
      $|J|$ and $|V^{MNS}_{e3}|$
      for $\tau$ leptogenesis model type I(a) with inverted hierarchy
      and $X_1 \le X_2$.
    }
  \label{Ve3J}
\end{figure}

To clarify the differences of predictions between inverted
hierarchical case and normal hierarchical case, we have plotted
$|V^{MNS}_{e3}|$ versus $|J|$ in Figs. \ref{Ve3J1}-\ref{Ve3J}. When
$|V^{MNS}_{e3}| \ll 1$, $J$ is approximately proportional to
$|V^{MNS}_{e3}|$. By choosing the standard parametrization of MNS
matrix, we obtain, \bea J=(1-s_{13}^2) s_{13} c_{12} s_{12} c_{23}
s_{23} \sin \delta, \eea with $V^{MNS}_{e3}=s_{13} \exp(-i
\delta)$.
In Figs. \ref{Ve3J1}-\ref{Ve3J}, within good approximation,
we can find the linear correlation between $|J|$ and
$|V^{MNS}_{e3}|$. One can read $|\sin \delta|$ from the slope
since
\bea |\sin \delta| \simeq \frac{1}{c_{12} s_{12} c_{23} s_{23}}
\frac{|J|}{s_{13}}. \eea In type I models with normal
hierachy, $\mu-$ and $\tau-$leptogenesis are allowed. The allowed
range for $\sin \delta$ is, \bea 0 \le |\sin \delta| \lesssim 1.
\eea For type I with inverted hierachy, $ \sin \delta $ is almost
maximal, \bea |\sin \delta| \simeq 1, \eea which implies that CP
violating phase $\gamma$ takes some non-vanishing definite value.
By fitting the data of neutrino mixings, we have determined the
allowed ranges for the parameters which are presented in table
\ref{fit}.
\begin{table}
\caption{The parameters which are determined by fitting with mixing angles.
The values in parentheses are obtained by fitting with the magnitudes
of  MNS elements based on $3 \sigma $ based fit taken from
\cite{GG}. The others correspond
to $90 \%$ CL level fit in \cite{GG}.}
\begin{tabular}{|c|c|c|c|c|} \hline
 & & & $|\gamma|$ & $ \beta $ \\ \hline
II(a) $e$ inv.
 $X_1 \leq X_2$
& $(0.30) \le |u_{\mu 2}|^2 \le (0.67)$
& $(0.33) \le |u_{\tau 2}|^2 \le (0.70)$
& $ (1.4) \sim (1.8) $
& $ (1.5) $ \\
II(b) $e$ inv. $X_1 \geq X_2$ &
$(0.30) \le |u_{\mu 1}|^2 \le (0.67)$ &
$(0.33) \le |u_{\tau 1}|^2 \le (0.70)$ &
$(1.4) \sim (1.8) $ &
$(1.5)$   \\ \hline
I(a) $\mu$
nor. $X_1 \leq X_2$ &
$0.085 \le |u_{e 1}|^2 \le 0.29$ &
$0.24 \le |u_{\tau 2}|^2 \le 0.68$ &
$0 \sim 3.1$ &
$0.60 \sim  1.1$ \\
& $(0.050) \le |u_{e 1}|^2 \le (0.37)$ &
$(0.16) \le |u_{\tau 2}|^2 \le (0.75)$ &
$(0) \sim (3.1)$ &
$(0.47) \sim (1.1)$ \\ \hline
I(a) $\mu$ inv. $X_1 \leq X_2$ &
$(0.97) \le |u_{e 1}|^2 \le (1.0)$ &
 $(0.40) \le |u_{\tau 2}|^2 \le(0.68)$ &
 $(1.4) \sim (1.8)$ &
 $(1.5)$  \\ \hline
I(b) $\mu$ nor. $X_1 \geq X_2$
& $0.25 \le |u_{\tau 1}|^2 \le 0.68 $ &
 $0.082 \le |u_{e 2}|^2 \le 0.29 $ &
 $0 \sim 3.1$ & $ 0.60 \sim 1.1$  \\
& $(0.16) \le |u_{\tau 1}|^2 \le (0.75)$ &
 $(0.050) \le |u_{e 2}|^2 \le (0.37)$ &
 $(0) \sim (3.1)$ & $ (0.47) \sim (1.1)$ \\ \hline
I(b) $\mu$ inv.$X_1 \geq X_2$ &
$(0.33) \le |u_{\tau 1}|^2 \le (0.70)$ &
 $(0.97) \le |u_{e 2}|^2 \le (1.0)$ &
 $(1.4) \sim (1.8)$ & $(1.5)$ \\ \hline
I(a) $\tau$ nor. $X_1 \leq X_2$ &
$0.093 \le |u_{e 1}|^2 \le 0.29$ &
 $0.28 \le |u_{\mu 2}|^2 \le 0.71$ &
 $0 \sim 3.1$ &
 $0.63 \sim 1.1 $  \\
 &
$(0.054) \le |u_{e 1}|^2 \le (0.38)$ &
 $(0.18) \le |u_{\mu 2}|^2 \le (0.78)$ &
 $(0) \sim (3.1)$ &
 $(0.50) \sim (1.2)$  \\ \hline
I(a) $\tau$ inv. $X_1 \leq X_2$ &
$(0.98) \le |u_{e 1}|^2 \le (1.0)$ &
 $(0.30) \le |u_{\mu 2}|^2 \le(0.64)$ &
 $(1.4) \sim (1.8)$ &
 $(1.5)$  \\ \hline
I(b) $\tau$ nor. $X_1 \geq X_2$
& $0.28 \le |u_{\mu 1}|^2 \le 0.71 $ &
 $0.092 \le |u_{e 2}|^2 \le 0.29 $ &
 $0 \sim 3.1$ & $ 0.63 \sim 1.1$  \\
& $(0.18) \le |u_{\mu 1}|^2 \le (0.78)$ &
 $(0.054) \le |u_{e 2}|^2 \le (0.37)$ &
 $(0) \sim (3.1)$ & $ (0.50) \sim (1.2)$ \\ \hline
I(b) $\tau$ inv.$X_1 \geq X_2$ &
$(0.30) \le |u_{\mu 1}|^2 \le (0.67)$ &
 $(0.97) \le |u_{e 2}|^2 \le (1.0)$ &
 $(1.4) \sim (1.8)$ & $(1.5)$ \\ \hline
\end{tabular}
\label{fit}
\end{table}
\\
\subsection{Unitarity triangle}
\begin{figure}
\setlength{\unitlength}{1mm}
\begin{picture}(100,40)
\put(96,15){\makebox(20,10)[r]{$V_{e3} V^{\ast}_{\mu3}=s_{13} c_{13} s_{23} \exp(-i \delta)$}}
\put(29,5){\makebox(20,10)[r]{$V_{e1} V^{\ast}_{\mu1}$}}
\put(40,30){\makebox(20,10)[r]{$V_{e2} V^{\ast}_{\mu2}$}}
\put(40,-10){\makebox(10,10)[r]{-$c_{13} s_{12} c_{12} c_{23}$}}
\put(70,0){\vector(-1,0){70}}
\put(70,0){\vector(-2,1){50}}
\put(0,0){\line(3,4){30}}
\put(30,41){\line(1,0){70}}
\put(19.5,25){\line(5,1){80}}
\put(100,41){\vector(-3,-4){30}}
\put(0,0){\makebox(-2,5)[r]{O}}
\put(19,25){\makebox(-2,5)[r]{B}}
\put(30,41){\makebox(-2,5)[r]{A}}
\end{picture}
\caption{Schematic view of unitarity triangle}
\label{tri}
\end{figure}
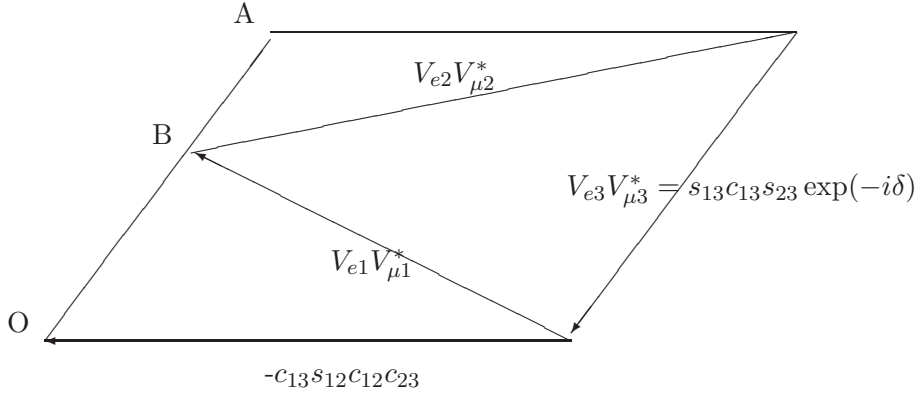

\begin{figure}
  \includegraphics{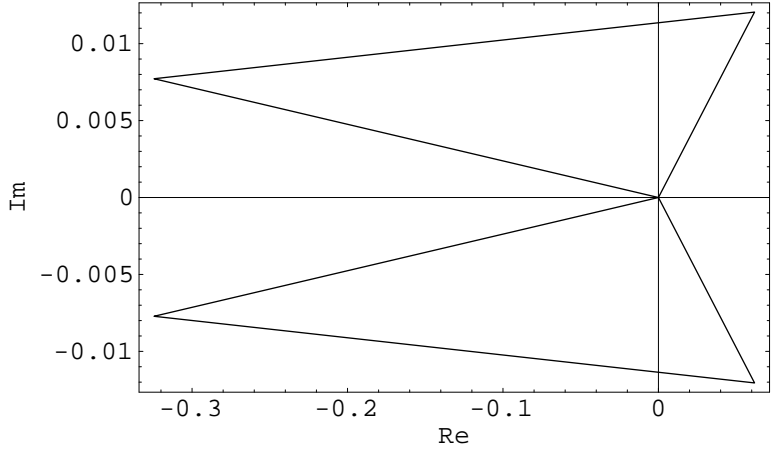}
    \caption{
      Unitarity triangles
      for $\tau$ leptogenesis model type I(a)
      with normal hierarchy which correspond to
      $
|V_{e 1}| \simeq 0.80, |V_{e 2}| \simeq 0.60, |V_{e 3}| \simeq 0.098,
|V_{\mu 1}| \simeq 0.41, |V_{\mu 2}| \simeq 0.65, |V_{\mu 3}| \simeq 0.64,
|V_{\tau 3}| \simeq 0.76,
|J| \simeq 0.0044$, $|\sin \delta| \simeq 0.19$, $|\gamma| \simeq 3.0 $
and $ \beta \simeq 0.94 $.
    }
  \label{triaN}
\end{figure}

\begin{figure}
  \includegraphics{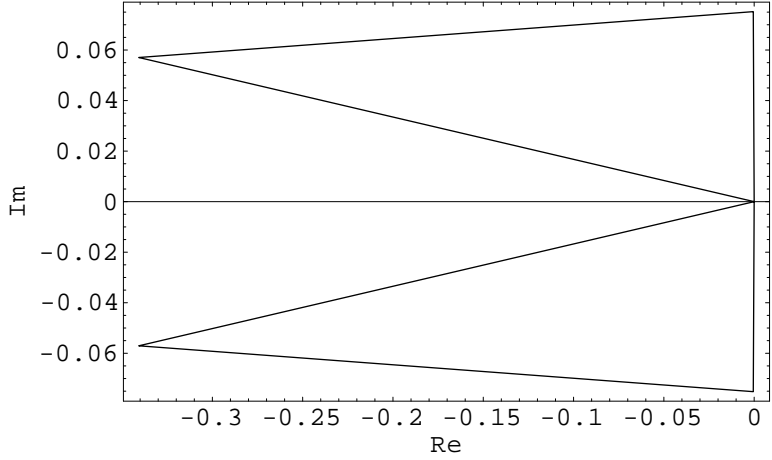}
    \caption{
      Unitarity triangles
      for $\tau$ leptogenesis model type I(a)
      inverted hierarchy which correspond to
 $|V_{e 1}| \simeq 0.86, |V_{e 2}| \simeq 0.49, |V_{e 3}| \simeq 0.13,
|V_{\mu 1}| \simeq 0.40, |V_{\mu 2}| \simeq 0.70, |V_{\mu 3}| \simeq 0.59, 
|V_{\tau 3}| \simeq 0.80,
|J| \simeq 0.026$, $ |\sin \delta| \simeq 1.0, |\gamma| \simeq 1.6 $
and $ \beta \simeq 1.5 $.}
  \label{triaI}
\end{figure}

Further one can reconstruct the unitarity triangles of the models
with two zeros texture which can satisfy the experimental
constraints. We focus on the unitarity triangle of $\mu-e$ sector
\bea && V^{MNS}_{e1} V^{MNS \ast}_{\mu 1} + V^{MNS}_{e2} V^{MNS
\ast}_{\mu 2}
+ V^{MNS}_{e3} V^{MNS \ast}_{\mu 3}=0, \nn \\
&& V^{MNS}_{e1} V^{MNS \ast}_{\mu 1}=-c_{13} \left(c_{12}
s_{12}c_{23}
+c_{12}^2 s_{23} s_{13} \exp(-i \delta) \right), \nn \\
&& V^{MNS}_{e2} V^{MNS \ast}_{\mu 2}=  c_{13}
\left(s_{12}c_{12}c_{23}-s_{12}^2 s_{23}s_{13} \exp(-i \delta) \right) ,\nn \\
&& V^{MNS}_{e3} V^{MNS \ast}_{\mu 3}= +c_{13} s_{13} s_{23}
\exp(-i\delta). \eea First we show the triangle schematically in
Fig. \ref{tri}. The triangle can be drawn inside a parallelogram
as shown in Fig. \ref{tri}. We note that, \bea \rm{OB:AB} =
c_{12}^2:s_{12}^2, \eea and $\delta$ is argument between
$V_{e3}^{MNS}$ and real axis.
In Fig.\ref{triaN} and in Fig.\ref{triaI}, we have shown the
triangle corresponding to the type I(a) $\tau$ leptogenesis for
normal and inverted hierarchical case, respectively. As we have
already noted, the inverted hierachical case, $\sin \delta$ is
almost maximal. Therefore the argument of $V_{e3}$ with respect to
real axis is 90 degree. For normal hierachical case, $|\sin
\delta|$ is smaller than $1$. Because only the magnitude of
$V^{MNS}$ is known, we have two fold ambiguities for $\delta$
even if the sizes of $s_{12}$, $s_{23}$,
$|V_{e3}^{MNS}|$ and $|J|$ are given. In Fig.
\ref{triaN}, we plot two triangles which correspond to
 $\delta$ and $-\delta$.
Two triangles which are related to each other by reflection with respect to
real axis can be distinguished by measuring the sign of $J$.
\section{Summary and Discussions}
In this work, we study CP violation in neutrino oscillations and
its possible connection with lepton family asymmetries
generated from heavy Majorana neutrino decays. We have derived a
general formula for CP violation in neutrino oscillations in
terms of heavy Majorana mass terms and Dirac mass terms. We
identify the two zeros texture models in which lepton asymmetry is
dominated by a particular family asymmetry. We have explored the
e-leptogenesis, $\mu-$leptogenesis and $\tau-$leptogenesis
scenarios and determined the allowed range of parameters from the
neutrino experimental results. Using the $ 90 \%$ and $ 3 \sigma $
bound on the magnitude of mixing angles measured at experiments,
we have constrained the parameters of the models. Based on the
analysis above, we have predicted the possible ranges of
$|V_{e3}^{MNS}|$ and the low energy CP violation observable $|J|$.
We have found that in the models with two zeros in $m_D$  and
inverted hierarchy, $|\sin \delta|$ is predicted to be almost
maximal.  Once those two unknown quantities are determined in
future neutrino oscillation experiments, we could compare them
with our predictions. Because the sign of J would be
determined from the measurment of CP violation via neutrino
oscillations, we can conclude whether the sign of CP violation at
low energy is consistent with  CP violation required in cosmology
\cite{BMNR} \cite{FGY} \cite{Endoh}.
\section{Acknowledgement}
We thank Z. Xiong and Z. Xing  for discussions. We also thank the
Yukawa Institute for Theoretical Physics at Kyoto University,
where this work was initiated during the workshop YITP-W-04-08 on
$\lq\lq$Summer Institute 2004" and YITP-W-04-21 on $\lq\lq$CP violatioin
and matter and anti-matter asymmetry". This work is supported by
the kakenhi of MEXT, Japan, No. 16028213(T.M.). SKK is supported
in part by BK21 program of the Ministry of Education in Korea and
in part by KOSEF Grant No. R01-2003-000-10229-0.

\end{document}